\documentclass[12pt]{article}
\usepackage{epsfig,youngtab,pdflscape,amssymb}
\usepackage{fancybox,amsmath,color}

\def \Tr{\mbox{Tr\,}}

\newcommand{\be}{\begin{equation}}
\newcommand{\bea}{\begin{eqnarray}}

\newcommand{\ee}{\end{equation}}
\newcommand{\eea}{\end{eqnarray}}

\begin{document}

\makeatletter
\@addtoreset{equation}{section}
\makeatother
\renewcommand{\theequation}{\thesection.\arabic{equation}}

\vspace{1.8truecm}

\vspace{15pt}


{\LARGE{ 
\centerline{\bf Anomalous dimensions from boson lattice models } 
}}  

\vskip.5cm 

\thispagestyle{empty} 
\centerline{ {\large\bf Shaun de Carvalho${}^{b,}$\footnote{{\tt 542425@students.wits.ac.za}}, 
Robert de Mello Koch$^{a,b,}$\footnote{{\tt robert@neo.phys.wits.ac.za}} }}

\centerline{ {\large\bf and Augustine Larweh Mahu${}^{b,}$\footnote{ {\tt a.larweh@aims.edu.gh}}}}

\vspace{.4cm}
\centerline{{\it ${}^a$ School of Physics and Telecommunication Engineering},}
\centerline{{ \it South China Normal University, Guangzhou 510006, China}}

\vspace{.4cm}
\centerline{{\it ${}^b$ National Institute for Theoretical Physics,}}
\centerline{{\it School of Physics and Mandelstam Institute for Theoretical Physics,}}
\centerline{{\it University of the Witwatersrand, Wits, 2050, } }
\centerline{{\it South Africa } }

\vspace{1truecm}

\thispagestyle{empty}

\centerline{\bf ABSTRACT}

\vskip.2cm 

Operators dual to strings attached to giant graviton branes in AdS$_5\times$S$^5$ can be described rather explicitly in the dual ${\cal N} = 4$ super Yang-Mills theory.
They have a bare dimension of order $N$ so that for these operators the large $N$ limit and the planar limit are distinct: summing only the planar diagrams will not capture the large $N$ dynamics. 
Focusing on the one-loop $SU(3)$ sector of the theory, we consider operators that are a small deformation of a 
${1\over 2}-$BPS multi-giant graviton state. 
The diagonalization of the dilatation operator at one loop has been carried out, but explicit formulas for the operators of
a good scaling dimension are only known when certain terms which were argued to be small, are neglected. 
In this article we include the terms which were neglected. 
The diagonalization is achieved by a novel mapping which replaces the problem of diagonalizing the dilatation operator with a system of bosons hopping on a lattice. 
The giant gravitons define the sites of this lattice and the open strings stretching between distinct giant gravitons define the
hopping terms of the Hamiltonian. 
Using the lattice boson model, we argue that the lowest energy giant graviton states are obtained by distributing the
momenta carried by the $X$ and $Y$ fields evenly between the giants with the condition that any particular giant carries only $X$ or $Y$ momenta, but not both.

\setcounter{page}{0}
\setcounter{tocdepth}{2}

\newpage
\tableofcontents

\setcounter{footnote}{0}

\linespread{1.1}
\parskip 4pt

{}~
{}~

\section{ Introduction }

Motivated by the AdS/CFT correspondence\cite{Maldacena:1997re,Gubser:1998bc,Witten:1998qj}, there has been 
dramatic progress in computing the planar spectrum of anomalous dimensions in ${\cal N} =4$ super Yang-Mills theory. 
The planar spectrum is now known, in principle, to all orders in the 't Hooft coupling \cite{Gromov:2013pga}. 
This has been possible thanks to the discovery of integrability\cite{Minahan:2002ve,Beisert:2010jr} in the planar limit 
of the theory. 
This spectrum of anomalous dimensions reproduces classical string energies on the
AdS$_5\times$S$^5$ spacetime, in the dual string theory\cite{Beisert:2005bm}.

Much less is known about ${\cal N} = 4$ super Yang-Mills theory outside the planar limit. 
There are many distinct large N but non-planar limits of the theory that could be considered and these correspond 
to a variety of fascinating physical problems. 
For example, the problem of considering new spacetime geometries (including black hole solutions) corresponds to 
considering operators with a bare dimension of order $N^2$\cite{Lin:2004nb}, while giant graviton 
branes\cite{McGreevy:2000cw,Grisaru:2000zn,Hashimoto:2000zp} are dual to operators with a bare dimension of 
order $N$. 
The planar limit does not correctly capture the dynamics of these operators\cite{Balasubramanian:2001nh,Berenstein:2003ah}.

Although much less is known about these large $N$ but non-planar limits, some progress has been made. 
Approaches based on group representation theory provide a powerful tool, essentially because they allow us to 
map the problem of the dynamics of the non-planar limit - summing the ribbon graphs contributing to correlation 
functions - into a purely algebraic problem in group theory. 
Typically, it can be phrased as the construction of a collection of projection operators and their properties. 
Once the algebraic problem is properly formulated, systematic approaches to it can be developed. 
As an example of this approach, bases of local gauge invariant operators have been
given\cite{Corley:2001zk,deMelloKoch:2007rqf,Kimura:2007wy,Brown:2007xh,Bhattacharyya:2008rb,Brown:2008ij,Kimura:2008ac,Kimura:2012hp}. 
These bases provide a good starting point from which the anomalous dimensions can be studied. 
This is basically because they diagonalize the free field two point function and, at weak coupling, operator mixing is highly
constrained\cite{deMelloKoch:2007nbd,Bekker:2007ea,Brown:2008rs,Koch:2010gp,DeComarmond:2010ie}. 
The resulting operators have a complicated multi-trace structure, quite different to the single trace structure relevant 
for the planar limit and its mapping to an integrable spin chain. 
The spectrum of anomalous dimensions has been computed for operators that are small deformations of 1/2 BPS operators.
Problems with 2 distinct characters have been solved: It is possible to simply treat all fields in the operator on the same 
footing, construct the basis and then 
diagonalize \cite{Carlson:2011hy,Koch:2011hb,deMelloKoch:2011ci,deMelloKoch:2012ck} or alternatively, one can 
build operators that realize a spacetime geometry or a giant graviton brane and use words constructed
from the fields of the CFT to describe string 
excitations\cite{deMelloKoch:2007nbd,Koch:2015pga,Koch:2016jnm}.
In the approach that treats all fields on the same footing, one simply defines the operators of the basis and considers 
the diagonalization of the dilatation operator with no physical input from the dual gravity description. 
When considering states dual to systems of giant gravitons, the Gauss Law of the dual giant world volume gauge 
theory emerges, so that in this approach we see open string and membranes are present in the CFT Hilbert space. 
When using words to describe string excitations, computations in the CFT reproduce the classical values of energies 
computed in string theory\cite{Koch:2015pga,Koch:2016jnm}, the worldsheet S-matrix\cite{Beisert:2005tm} and 
has lead to the discovery of integrable subsectors for closed string excitations of certain LLM 
backgrounds\cite{Koch:2016jnm}. 
Clearly, this is a rich problem with hidden simplicity, so that further study of these limits are bound to be fruitful. 
The existence of this hidden simplicity is not unexpected: conventional lore of the large $N$ limit identifies $1/N$ 
as the gravitational interaction, so that the $N\to\infty$ limit, in which this interaction
is turned off, should be a simple limit.

One next step that can be contemplated, is to go beyond small perturbations of the 1/2 BPS sector. 
This problem is our main motivation in this study, and we will take a small step in this direction. 
We will study operators constructed from three complex adjoint scalars $X$, $Y$, $Z$ of ${\cal N} = 4$ 
super Yang-Mills theory. 
Operators that are a small perturbation of a 1/2 BPS operator are constructed using mainly $Z$ fields. 
For these operators, interactions between the $X,Y$ fields are subdominant to interactions between 
$X,Z$ and between $Y,Z$ fields and can hence be neglected. As we move further
from the original 1/2 BPS operator, more and more $X,Y$ fields are added.
At some point the interactions between the $X,Y$ fields can no longer be neglected.
Dealing with these interactions is the focus of our study. 
We will argue that this is a well defined problem, that can be solved, often explicitly.
This is accomplished by phrasing the new $X,Y$ interactions as a lattice model, for essentially free bosons. 
Thus, we finally land up with a simple problem that is familiar and can be solved. 
This is the basic achievement of this paper.

Our results show a fascinating structure that deserves to be discussed.
The mapping to the lattice model associates a harmonic oscillator to both the $X$ field and to the $Y$ field. 
Earlier results \cite{deMelloKoch:2011ci} treating the leading term, performed the diagonalization by associating a 
harmonic oscillator to the $Z$ field, so that in the end we seem to be seeing an equality in the description of
the three scalar fields. 
An even-handed treatment of all three fields is a big step towards being able to treat operators constructed with 
equal numbers of $X$, $Y$ and $Z$ fields. 
This would most certainly go beyond the ${1\over 2}$-BPS sector, the main motivation for our study.

In the next section we review the action of the one loop dilatation operator $D_2$. 
The action of $D_2$ in the $SU(3)$ sector, in the Schur polynomial basis, has been evaluated 
previously\cite{Koch:2013yaa} and we simply quote and use the result.
We then move to the Gauss graph basis of \cite{deMelloKoch:2012ck}, in which the terms in $D_2$ arising from 
$Z,Y$ or $Z,X$ interactions are diagonal. 
Again, this is a known result and we simply use it. 
The Gauss graph basis has a natural interpretation in terms of giant graviton branes and their open string excitations. 
We will often use this language of branes and strings. 
We then come to the central term of interest: the term in $D_2$ arising from $X,Y$ interactions. 
Denote this term by $D^{XY}_2$. 
We will carefully evaluate this term, arriving at a rather simple formula, which is the starting point for section \ref{blattice}. 
The explicit expression for $D^{XY}_2$ can easily be identified with a lattice model for a collection of bosons. 
The giant gravitons define the sites of this lattice, and the open string excitations determine the lattice Hamiltonian. 
Section \ref{diag} diagonalizes the dilatation operator for a number of giants plus open string configurations,
arriving at detailed and explicit expressions both for the anomalous dimensions and for the operators of a definite
scaling dimension. 
Our conclusions and some discussion are given in section \ref{conc}.

\section{Action of the One Loop Dilatation Operator}

We combine the 6 hermitian adjoint scalars of ${\cal N} = 4$ super Yang-Mills theory into three complex combinations,
denoted $X,Y,Z$. 
The operators we consider are constructed using $n$ $Z$s, $m$ $Y$s and $p$ $X$s. 
Operators that are dual to giant graviton branes are constructed using $n+m+p\sim N$ fields. 
We will focus on operators that are small deformations of 1/2 BPS operators,
achieved by choosing $n\gg m+ p$. We will fix ${m\over p}\sim 1$ as $N\to\infty$ and treat ${m\over n}$
as a small parameter. 
The collection of operators constructed using $X, Y, Z$ fields are often referred to as the $SU(3)$ sector. 
This is not strictly speaking correct since these operators do mix with operators containing fermions. 
At one loop however, this is a closed sector.

Our starting point is the action of the one loop dilatation operator of the $SU(3)$ sector\
\bea
D_2 = D^{YZ}_2 + D^{XZ}_2 + D^{XY}_2 
\eea
where
\bea
D^{AB}_2 \equiv g^2_{YM}\Tr \left([A,B][\partial_A, \partial_B]\right)
\eea
on the restricted Schur polynomial basis. This has been evaluated in \cite{Koch:2013yaa}.
Further, the terms $D^{YZ}_2$ and $D^{XZ}_2$ have been diagonalized. 
The operators of a definite scaling dimension $O_{R,r}(\sigma)$, called Gauss graph 
operators\cite{Koch:2011hb,deMelloKoch:2012ck}, are labeled by a pair of Young diagrams 
$R\vdash n +m+ p$ and $r \vdash n$ as well as a permutation $\sigma\in  S_m \times S_p$. 
Although these labels arise when diagonalizing $D^{YZ}_2$ and $D^{XZ}_2$ in the CFT, they have a natural 
interpretation in the dual gravitational description in terms of giant graviton branes plus open string
excitations. 
A Young diagram $R$ that has $q$ rows corresponds to a system of $q$ giant gravitons. 
The $Y$ and $X$ fields describe the open string excitations of these giants, so that there are $m + p$ open strings in total. 
We can describe the state of the system using a graph, with nodes of the graph representing the branes (and hence rows of 
$R$) and directed edges of the graph describing the open string excitations (represented by $X$ and $Y$ fields in the CFT).
Each directed edge ends on any two (not necessarily distinct) of the $q$ branes. 
The only configurations that appear when $D^{YZ}_2$ and $D^{XZ}_2$ are diagonalized have the same number of 
strings starting or terminating on any given giant, for the $X$ and $Y$ strings 
separately\cite{deMelloKoch:2012ck,Koch:2013yaa}. 
Thus the Gauss Law of the brane world volume theory implied by the fact that the giant graviton has a compact
world volume\cite{Balasubramanian:2004nb} emerges rather naturally in the CFT description. 
Since every terminating edge endpoint can be associated to a unique emanating endpoint, we can give a nice description 
of how the open strings are connected to the giants by specifying how the terminating and emanating endpoints
are associated. 
The permutation $\sigma\in S_m\times S_p$ describes how the $m$ $Y$'s and the $p$ $X$'s are draped between the 
$q$ giant gravitons by describing this association\cite{deMelloKoch:2012ck,Koch:2013yaa}. The explicit form of the 
Gauss graph operators is\cite{deMelloKoch:2012ck,Koch:2013yaa}
\bea
\label{ggops}
O^{\vec m,\vec p}_{R,r}(\sigma)=
{|H_X\times H_Y|\over\sqrt{p!m!}}\sum_{j,k}\sum_{s\vdash m}\sum_{t\vdash p}\sum_{\vec\mu_1,\vec\mu_2}
\sqrt{d_s d_t}\Gamma^{(s,t)}_{jk}(\sigma)\cr
\times B_{j\vec\mu_1}^{(s,t)\to 1_{H_X\times H_Y}}
B_{k\vec\mu_2}^{(s,t)\to 1_{H_X\times H_Y}}
O_{R,(t,s,r)\vec\mu_1\vec\mu_2}
\eea
Each box in $R$ is associated with one of the complex fields. 
$r$ is a label for the $Z$ fields. 
The graph $\sigma$ encodes important information. 
The number of $Y$ (or $X$) strings terminating on the $i$th node which equals the number of
$Y$ (or $X$) strings emanating from the $i$th node is denoted by $m_i$ (or $p_i$). $m_i$ (or $p_i$) also counts the 
number of boxes in the $i$th row of $R$ that correspond to $Y$ (or $X$) fields. We will often assemble $m_i$ and $p_i$ 
into the vectors $\vec m$ and $\vec p$. 
The number of $Y$ (or $X$) strings stretching between nodes $i$ and $k$ is denoted $m_{ik}$ (or $p_{ik}$), 
while the number of strings stretching from node $i$ to node $k$ are denoted $m_{i\to k}$ (or $p_{i\to k}$). 
A Young diagram with $k$ boxes $a\vdash k$ labels an irreducible representation of $S_k$ with dimension $d_a$. 
The branching coefficients $B_{j\vec\mu_1}^{(s,t)\to 1_{H_X\times H_Y}}$ resolve the operator that projects from 
$(s,t)$, with $s \vdash m$, $t \vdash p$, an irreducible representation of $S_m\times S_p$, to the trivial (identity)
representation of the product group $H_Y\times H_X$ with $H_Y = S_{m_1}\times S_{m_2}\times \cdots S_{m_q}$
and $H_X = S_{p_1}\times S_{p_2}\times \cdots S_{p_q}$ , i.e.
\bea
{1\over H_X \times H_Y}\sum_{\gamma\in H_X\times H_Y}
\Gamma^{(s,t)}_{ik} (\gamma) =\sum_{\vec{\mu}}
B_{i\vec\mu}^{(s,t)\to 1_{H_X\times H_Y}}B_{k\vec\mu}^{(s,t)\to 1_{H_X\times H_Y}}
\eea
$\Gamma^{(s,t)}_{jk}(\sigma)$ is a matrix (with row and column indices $jk$) representing $\sigma\in S_m\times S_p$
in irreducible representation $(s,t)$. The operators $O_{R,(r,s,t)\vec\mu_1\vec\mu_2}$ are
normalized versions of the restricted Schur polynomials \cite{Bhattacharyya:2008rb}
\bea
\chi_{R,(t,s,r)\vec\mu_1\vec\mu_2}(Z,Y,X)=
{1\over n!m!p!}\sum_{\sigma\in S_{n+m+p}}
\chi_{R,(t,s,r)\vec\mu_1\vec\mu_2}(\sigma)\Tr (\sigma Z^{\otimes n}Y^{\otimes m}X^{\otimes p})\cr
\eea
which themselves provide a basis for the gauge invariant operators of the theory. 
The restricted characters $\chi_{R,(t,s,r)\vec\mu_1\vec\mu_2}(\sigma)$ are defined by tracing the
matrix representing group element $\sigma$ in representation $R$ over the subspace
giving an irreducible representation $(r,s,t)$ of the $S_n\times S_m\times S_p$ subgroup.
There is more than one choice for this subspace and the multiplicity labels $\vec\mu_1\vec\mu_2$ resolve this ambiguity, 
for the row and column index of the trace. 
The operators $O_{R,(t,s,r)\vec\mu_1\vec\mu_2}$ given by
\bea
O_{R,(t,s,r)\vec\mu_1\vec\mu_2}=\sqrt{{\rm hooks}_r{\rm hooks}_s{\rm hooks}_t\over {\rm hooks}_R f_R}
\chi_{R,(t,s,r)\vec\mu_1\vec\mu_2}
\eea
have unit two point function. hooks$_r$ stands for the product of hook lengths of Young diagram $r$ and $f_R$ 
stands for the product of the factors of Young diagram $R$. 
The action of the dilatation operator on the Gauss graph operators is \cite{Koch:2011hb,deMelloKoch:2012ck,Koch:2013yaa}
\bea
\label{ggaction}
D^{YZ}_2 O^{\vec m,\vec p}_{R,r}(\sigma)
=-g_{YM}^2\sum_{i<j}m_{ij}(\sigma)\Delta_{ij}O^{\vec m,\vec p}_{R,r}(\sigma)\cr
D^{XZ}_2 O^{\vec m,\vec p}_{R,r}(\sigma)
=-g_{YM}^2\sum_{i<j}p_{ij}(\sigma)\Delta_{ij}O^{\vec m,\vec p}_{R,r}(\sigma)
\eea
where $\Delta_{ij} = \Delta_{ij}^-+\Delta_{ij}^0+\Delta_{ij}^+$\cite{deMelloKoch:2011ci}. 
We will now spell out the action of the operators $\Delta_{ij}^+$, $\Delta_{ij}^0$ and $\Delta_{ij}^-$. 
Denote the row lengths of $r$ by $l_{r_i}$. 
The Young diagram $r^+_{ij}$ is obtained by deleting a box from row $j$ and adding it to row $i$.
The Young diagram $r^-_{ij}$ is obtained by deleting a box from row $i$ and adding it to row $j$.
In terms of these Young diagrams we have
\bea
\Delta_{ij}^0 O^{\vec m,\vec p}_{R,r}(\sigma) = -(2N+l_{r_i}+l_{r_j})O^{\vec m,\vec p}_{R,r}(\sigma)
\eea
\bea
\Delta_{ij}^+ O^{\vec m,\vec p}_{R,r}(\sigma) = \sqrt{(N+l_{r_i})(N+l_{r_j})}
O^{\vec m,\vec p}_{R^+_{ij},r^+_{ij}}(\sigma)
\eea
\bea
\Delta_{ij}^- O^{\vec m,\vec p}_{R,r}(\sigma) = \sqrt{(N+l_{r_i})(N+l_{r_j})}
O^{\vec m,\vec p}_{R^-_{ij},r^-_{ij}}(\sigma)
\eea
Notice that $D^{YZ}_2$ and $D^{XZ}_2$ in (\ref{ggaction}) are not yet diagonal: they still mix operators with different 
$R,r$ labels. 
This last diagonalization however, is rather simple: it maps into diagonalizing a collection of decoupled oscillators
as demonstrated in \cite{deMelloKoch:2011ci}. 
We will call these $Z$ oscillators, since they are associated to the $r$ label which organizes the $Z$ fields. 
It is clear that $D^{XY}_2$ does not act on the $r$ label so that in the end, the contribution from $D^{XY}_2$ simply 
shifts the ground state eigenvalue of the $Z$ oscillators.

We will now focus on the term $D^{XY}_2$. 
Recall that our operators are built with many more $Z$ fields, than $X$ or $Y$ fields ($n\gg p + m$). 
Since this term contains no derivatives with respect to $Z$ it is subleading (of order ${m\over n}$) when
compared to $D^{YZ}_2$ and $D^{XZ}_2$. 
Diagonalizing this operator is the main goal of this article, so it is useful to sketch the derivation of the matrix elements 
of $D^{XY}_2$ in the Gauss graph basis. 
We will simply quote existing results that we need, giving complete details only for the final stages of the evaluation,
which are novel. 
The reader will find useful background material in \cite{Koch:2013yaa}. 
The action of this term on the restricted Schur polynomial basis was computed in \cite{Koch:2013yaa}. The result is
\bea
&&D^{XY}_2O_{R,(t,s,r)\vec\mu\vec\nu}
=\sum_{R'}\sum_{T,(y,x,w)\vec\alpha\vec\beta}{\cal C}\cr
&&\Tr _{R\oplus T}\left(\left[P_1,\Gamma^{R}(1,p+1)\right]I_{R'T'}\left[P_2,\Gamma^T(1,p+1)\right]I_{T',R'}\right)
O_{T,(y,x,w)\vec\beta\vec\alpha}\nonumber
\eea
where
\bea
{\cal C} &=& -g^2_{YM}c_{RR'}
{d_T m p\over d_x d_y d_w (n+m+p)d_{R'}}
\sqrt{f_T{\rm hooks}_T{\rm hooks}_r{\rm hooks}_s{\rm hooks}_t\over
f_R{\rm hooks}_R{\rm hooks}_w{\rm hooks}_x{\rm hooks}_y}\cr
P_1&=& P_{R,(t,s,r)\vec\mu\vec\nu}\qquad\qquad P_2 = P_{T,(y,x,w)\vec\alpha\vec\beta}
\eea
$\Gamma^S(\sigma)$ is the matrix representing $\sigma\in S_{n+m+p}$ in irreducible representation $S\vdash n+m+p$. 
Young diagram $R'$ is obtained from Young diagram $R$ by dropping a single box, with $c_{RR'}$ denoting the factor 
of this box. 
$I_{T'R'}$, $I_{R'T'}$, $P_1$ and $P_2$ are intertwining maps. 
$I_{T'R'}$ maps from the carrier space of $R'$ to the carrier space of $T'$. 
It is only non-vanishing if $T'$ and $R'$ are equal as Young diagrams implying that operators labeled by $R$ and $T$ 
can only mix if they differ by the placement of a single box. 
The operators $P_1$ and $P_2$ are the intertwining maps used in the construction of the restricted Schur polynomials. 
It is challenging to evaluate the above expression explicitly, basically because it is difficult to construct $P_1$ and $P_2$.
However, the above expression has not yet employed the simplifications of large $N$. 
To do this, following \cite{Koch:2011hb} we will use the displaced corners approximation. 
After applying the approximation we obtain\cite{Koch:2013yaa}
\bea
D^{XY}_2O_{R,(t,s,r)\vec\mu\vec\nu}
=\sum_{T,(w,v,u)\vec\alpha\vec\beta}\tilde M_{R,(t,s,r)\vec\mu\vec\nu\,\,T,(w,v,u)\vec\alpha\vec\beta}
O_{T,(w,v,u)\vec\alpha\vec\beta}
\eea
where
\bea
\label{dca}
\tilde M_{R,(t,s,r)\vec\mu\vec\nu\,\,T,(w,v,u)\vec\alpha\vec\beta}=-g_{YM}^2\sum_{R'}
\delta_{R_i'T_k'}\delta_{ru}{pm\over\sqrt{d_s d_t d_w d_v}}\sqrt{c_{RR'}c_{TT'}\over l_{R_i}l_{T_k}}\times\cr
\Tr\Big[
E^{(1)}_{ki}P^{(\vec p,\vec m)}_{t\alpha_1\beta_1;s\alpha_2\beta_2}
E^{(p+1)}_{ik}P^{(\vec p',\vec m')}_{w\mu_1\nu_1;v\mu_2\nu_2}
-E^{(1)}_{ci}E^{(p+1)}_{kc}P^{(\vec p,\vec m)}_{t\alpha_1\beta_1;s\alpha_2\beta_2}
E^{(1)}_{ak}E^{(p+1)}_{ia}P^{(\vec p',\vec m')}_{w\mu_1\nu_1;v\mu_2\nu_2}\cr
-E^{(1)}_{kc}E^{(p+1)}_{ci}P^{(\vec p,\vec m)}_{t\alpha_1\beta_1;s\alpha_2\beta_2}
E^{(1)}_{ia}E^{(p+1)}_{ak}P^{(\vec p',\vec m')}_{w\mu_1\nu_1;v\mu_2\nu_2}
+E^{(p+1)}_{ki}P^{(\vec p,\vec m)}_{t\alpha_1\beta_1;s\alpha_2\beta_2}
E^{(1)}_{ik}P^{(\vec p',\vec m')}_{w\mu_1\nu_1;v\mu_2\nu_2}
\Big]\cr
\eea
The trace in this expression is over the tensor product $V^{\otimes n+m}_p$ where $V_p$ is the fundamental 
representation of $U(p)$. 
The intertwining maps used to define the restricted Schur polynomials ($P_1$ and $P_2$ above) factor into an action on
the boxes associated to the $Z$ fields, an action on the boxes associated to the $Y$ fields and an action on the boxes
associated to the X fields. The intertwining maps\footnote{A very explicit algorithm for the construction of these maps 
has been given in \cite{Koch:2011hb}.} $P^{(\vec p,\vec m)}_{t\alpha_1\beta_1;s\alpha_2\beta_2}$ and
$P^{(\vec p',\vec m')}_{w\mu_1\nu_1;v\mu_2\nu_2}$ are the actions of the intertwining maps on the $X$ and $Y$ 
fields only. 
This happens because the trace over the Z field indices, which is simple as the dilatation operator $D^{XY}_2$ does not 
act on the Z fields, has been performed. Young diagram $R'_i$ is obtained from $R$ by dropping a single box from row 
$i$ and $T'_k$ from $T$ by dropping a single box from row $k$.

The result (\ref{dca}) gives the $D^{XY}_2$ term in the dilatation operator, as a matrix that must be diagonalized. 
As we will see, all three terms in $D_2$ are simultaneously diagonalizable at large $N$ so that it is convenient to employ
the Gauss graph basis which already diagonalizes both $D^{ZY}_2$ and $D^{ZX}_2$. 
The problem of diagonalizing $D^{XY}_2$ then amounts to a diagonalization on degenerate subspaces of $D^{ZY}_2$ 
and $D^{ZX}_2$. 
Thus, the original diagonalization of an enormous matrix is replaced by diagonalizing a number of smaller matrices - a 
significant simplification. Applying the results of \cite{Koch:2013yaa}, we find that, after the change in basis
\bea
D^{XY}_2 \hat O^{\vec m,\vec p}_{R,r}(\sigma_1)
= M^{\vec m,\vec p}_{R,r,\sigma_1\,\,T,t,\sigma_2}
\hat O^{\vec m,\vec p}_{T,t}(\sigma_2)
\eea
where
\bea
\label{MinDA}
&&M^{\vec m,\vec p}_{R,r,\sigma_1\,\,T,t,\sigma_2}=-g_{YM}^2
{1\over\sqrt{|O_{R,r}^{\vec m,\vec p}(\sigma_1)|^2 |O_{T,t}^{\vec m,\vec p}(\sigma_2)|^2}}\times\cr\cr
&&\sum_{R'}{\delta_{R'_iT'_k}\delta_{ru}\over (p-1)!(m-1)!}
\sqrt{c_{RR'}c_{TT'}\over l_{R_i}l_{T_k}}
\sum_{\psi_1\in S_{\vec p}\times S_{\vec m}}\sum_{\psi_2\in S_{\vec p'}\times S_{\vec m'}}\cr\cr
&&\Big[
\langle\vec{p}',\vec{m}'|\sigma_2\psi_2^{-1}E^{(1)}_{ki}\psi_1|\vec p,\vec m\rangle
\langle \vec p,\vec m|\sigma_1^{-1}\psi_1^{-1}E^{(p+1)}_{ik}\psi_2 |\vec p',\vec m'\rangle\cr\cr
&&-\langle\vec{p}',\vec{m}'|\sigma_2\psi_2^{-1}E^{(1)}_{ci}E^{(p+1)}_{kc}\psi_1|\vec p,\vec m\rangle
\langle \vec p,\vec m|\sigma_1^{-1}\psi_1^{-1}E^{(1)}_{ak}E^{(p+1)}_{ia}\psi_2 |\vec p',\vec m'\rangle\cr\cr
&&-\langle\vec{p}',\vec{m}'|\sigma_2\psi_2^{-1}E^{(1)}_{kc}E^{(p+1)}_{ci}\psi_1|\vec p,\vec m\rangle
\langle \vec p,\vec m|\sigma_1^{-1}\psi_1^{-1}E^{(1)}_{ia}E^{(p+1)}_{ak}\psi_2 |\vec p',\vec m'\rangle\cr\cr
&&+\langle\vec{p}',\vec{m}'|\sigma_2\psi_2^{-1}E^{(p+1)}_{ki}\psi_1|\vec p,\vec m\rangle
\langle \vec p,\vec m|\sigma_1^{-1}\psi_1^{-1}E^{(1)}_{ik}\psi_2 |\vec p',\vec m'\rangle\Big]
\eea
Here the Gauss graph operators $\hat O^{\vec m,\vec p}_{R,r}(\sigma_1)$ are normalized to have a unit two
point function. 
They are related to the operators introduced in (\ref{ggops}) as follows
\bea
O_{R,r}^{\vec m,\vec p}(\sigma)
=\sqrt{\prod_{i=1}^q m_{ii}(\sigma)! m_{ii}(\sigma)!\prod_{k,l,k\ne l}m_{k\to l}(\sigma)!p_{k\to l}(\sigma)!}\,\,\,
\hat O_{R,r}^{\vec m,\vec p}(\sigma)
\eea
Introduce the vectors $(v^{(i)})_a = \delta_{ia}$ which form a basis for $V_p$. 
The vector $|\vec p,\vec m\rangle$ is defined as follows
\bea
|\vec p,\vec m\rangle = |\vec p\rangle\otimes |\vec m\rangle
\eea
where
\bea
|\vec p\rangle &=& (v^{(1)})^{\otimes p_1}\otimes\cdots\otimes (v^{(q)})^{\otimes p_q}\cr
|\vec m\rangle &=& (v^{(1)})^{\otimes m_1}\otimes\cdots\otimes (v^{(q)})^{\otimes m_q}
\eea

We will now explain how the sums over $\psi_1$ and $\psi_2$ in (\ref{MinDA}) can be evaluated.
This discussion is novel and is one of the new contributions of this paper. Consider the term
\bea
T_1 =\sum_{\psi_1\in S_{\vec p}\times S_{\vec m}}\sum_{\psi_2\in S_{\vec p'}\times S_{\vec m'}}
\langle\vec{p}',\vec{m}'|\sigma_2\psi_2^{-1}E^{(1)}_{ki}\psi_1|\vec p,\vec m\rangle
\langle \vec p,\vec m|\sigma_1^{-1}\psi_1^{-1}E^{(p+1)}_{ik}\psi_2 |\vec p',\vec m'\rangle
\nonumber
\eea
The dependence on the permutations $\sigma_1,\sigma_2$ can be simplified with the following
change of variables: replace $\psi_2$ with $\tilde\psi_2$ where
\bea
\tilde\psi_2=\psi_2\sigma_2^{-1}\qquad\Rightarrow\qquad \tilde\psi_2^{-1}=\sigma_2\psi_2^{-1}
\eea
After relabeling $\tilde\psi_2\to\psi_2$ and taking the transpose of the first factor which is a real number, we find
\bea
T_1 =\sum_{\psi_1\in S_{\vec p}\times S_{\vec m}}\sum_{\psi_2\in S_{\vec p'}\times S_{\vec m'}}
\langle\vec{p},\vec{m}|\psi_1^{-1}E^{(1)}_{ik}\psi_2|\vec p',\vec m'\rangle
\langle \vec p,\vec m|\sigma_1^{-1}\psi_1^{-1}E^{(p+1)}_{ik}\psi_2\sigma_2 |\vec p',\vec m'\rangle
\nonumber
\eea
If $i \ne k$, the matrix element $\langle\vec{p},\vec{m}|\psi_1^{-1}E^{(1)}_{ik}\psi_2|\vec p',\vec m'\rangle$
is only non-vanishing if $\vec p\ne \vec p'$ and $\vec m = \vec m'$, while the matrix element
$\langle \vec p,\vec m|\sigma_1^{-1}\psi_1^{-1}E^{(p+1)}_{ik}\psi_2\sigma_2 |\vec p',\vec m'\rangle$
is only non-vanishing if $\vec p = \vec p'$ and $\vec m \ne \vec m'$. Thus, $T_1$ vanishes for $i\ne k$.
Indicate this explicitly as follows
\bea
T_1 = \delta_{ik}\sum_{\psi_1,\psi_2\in S_{\vec p}\times S_{\vec m}}
\langle\vec{p},\vec{m}|\psi_1^{-1}E^{(1)}_{ii}\psi_2|\vec p,\vec m\rangle
\langle \vec p,\vec m|\sigma_1^{-1}\psi_1^{-1}E^{(p+1)}_{ii}\psi_2\sigma_2 |\vec p,\vec m\rangle
\nonumber
\eea
To simplify this expression further, note that $E^{(1)}_{ii}|\vec p,\vec m\rangle$ is only non-zero if vector $v^{(i)}$ 
occupies slot one in the vector $|\vec p\rangle$. 
In this case $E^{(1)}_{ii}|\vec p,\vec m\rangle=|\vec p,\vec m\rangle$.
Since $\psi_1$ and $\psi_2$ shuffle the vectors in $|\vec p,\vec m\rangle$ into all possible locations, $E^{(1)}_{ii}$
will in the end count how many times the vector $v^{(i)}$ appears in $|\vec p,\vec m\rangle$. 
This is given by $p_i$ introduced above. 
A similar argument applies to $E^{(p+1)}|\vec p,\vec m\rangle$,
Thus, we obtain
\bea
T_1 &=& \delta_{ik}{p_i\over p}{m_i\over m}\sum_{\psi_1,\psi_2\in S_{\vec p}\times S_{\vec m}}
\langle\vec{p},\vec{m}|\psi_1^{-1}\psi_2|\vec p,\vec m\rangle
\langle \vec p,\vec m|\sigma_1^{-1}\psi_1^{-1}\psi_2\sigma_2 |\vec p,\vec m\rangle\cr\cr
&=& \delta_{ik}{p_i\over p}{m_i\over m}\sum_{\psi_1,\psi_2\in S_{\vec p}\times S_{\vec m}}
\sum_{h_1,h_2\in H_X\times H_Y}\delta (\psi_1^{-1}\psi_2 h_1)
\delta(\sigma_1^{-1}\psi_1^{-1}\psi_2\sigma_2 h_2)
\nonumber
\eea
Now, perform the following change of summation variables $\psi_1\to\tilde\psi_1$ with
\bea
\psi_1 = \psi_2 \tilde\psi_1
\eea

\noindent
The summand is now independent of $\psi_2$ so that after summing over $\psi_2$ and relabeling $\tilde\psi_1\to\psi_1$ 
we find
\bea
T_1 = \delta_{ik}(p-1)! (m-1)! p_i m_i \sum_{\psi_1\in S_{\vec p}\times S_{\vec m}}
\sum_{h_1,h_2\in H_X\times H_Y}\delta (\psi_1 h_1)
\delta(\sigma_1^{-1}\psi_1\sigma_2 h_2)\nonumber
\eea
Summing over $\psi_1$ now gives
\bea
T_1 = \delta_{ik}(p-1)! (m-1)! p_i m_i 
\sum_{h_1,h_2\in H_X\times H_Y}
\delta(\sigma_1^{-1}h_1^{-1}\sigma_2 h_2)
\eea

We also need to consider the term
\bea
T_4 &=&\sum_{\psi_1\in S_{\vec p}\times S_{\vec m}}\sum_{\psi_2\in S_{\vec p'}\times S_{\vec m'}}
\langle\vec{p}',\vec{m}'|\sigma_2\psi_2^{-1}E^{(p+1)}_{ki}\psi_1|\vec p,\vec m\rangle
\langle \vec p,\vec m|\sigma_1^{-1}\psi_1^{-1}E^{(1)}_{ik}\psi_2 |\vec p',\vec m'\rangle\cr
&=&\sum_{\psi_1\in S_{\vec p}\times S_{\vec m}}\sum_{\psi_2\in S_{\vec p'}\times S_{\vec m'}}
\langle \vec p,\vec m|\sigma_1^{-1}\psi_1^{-1}E^{(1)}_{ik}\psi_2 |\vec p',\vec m'\rangle
\langle\vec{p},\vec{m}|\psi_1^{-1}E^{(p+1)}_{ik}\psi_2\sigma_2^{-1}|\vec p',\vec m'\rangle\nonumber
\eea
Changing variables $\psi_1^{-1}\to\sigma_1^{-1}\psi_1^{-1}$ shows that $T_4 = T_1$ and hence
\bea
T_1 + T_4 = 2\delta_{ik}(p-1)! (m-1)! p_i m_i 
\sum_{h_1,h_2\in H_X\times H_Y}
\delta(\sigma_1^{-1}h_1^{-1}\sigma_2 h_2)
\eea

The next sum we consider is
\bea
T_2 =\sum_{\psi_1\in S_{\vec p}\times S_{\vec m}}\sum_{\psi_2\in S_{\vec p'}\times S_{\vec m'}}
\langle\vec{p}',\vec{m}'|\sigma_2\psi_2^{-1}E^{(1)}_{ci}E^{(p+1)}_{kc}\psi_1|\vec p,\vec m\rangle
\langle \vec p,\vec m|\sigma_1^{-1}\psi_1^{-1}E^{(1)}_{ak}E^{(p+1)}_{ia}\psi_2 |\vec p',\vec m'\rangle
\nonumber
\eea
Changing variables $\psi_2^{-1}\to\tilde\psi_2^{-1}$  with
\bea
\tilde\psi_2^{-1}=\sigma_2\psi_2^{-1}\qquad\Rightarrow\qquad\tilde\psi_2 =\psi_2\sigma_2^{-1}
\eea
the sum becomes
\bea
T_2 =\sum_{\psi_1\in S_{\vec p}\times S_{\vec m}}\sum_{\psi_2\in S_{\vec p'}\times S_{\vec m'}}
\langle\vec{p}',\vec{m}'|\psi_2^{-1}E^{(1)}_{ci}E^{(p+1)}_{kc}\psi_1|\vec p,\vec m\rangle
\langle \vec p,\vec m|\sigma_1^{-1}\psi_1^{-1}E^{(1)}_{ak}E^{(p+1)}_{ia}\psi_2\sigma_2 |\vec p',\vec m'\rangle\cr
=\sum_{\psi_1\in S_{\vec p}\times S_{\vec m}}\sum_{\psi_2\in S_{\vec p'}\times S_{\vec m'}}
\langle\vec{p}',\vec{m}'|\psi_2^{-1}\psi_1E^{\psi_1^{-1}(1)}_{ci}E^{\psi_1^{-1}(p+1)}_{kc}|\vec p,\vec m\rangle
\langle \vec p,\vec m|\sigma_1^{-1}E^{\psi_1^{-1}(1)}_{ak}E^{\psi_1^{-1}(p+1)}_{ia}
\psi_1^{-1}\psi_2\sigma_2 |\vec p',\vec m'\rangle
\nonumber
\eea
Change variables $\psi_2\to\rho$ with $\rho=\psi_1^{-1}\psi_2$ and relabel $\rho\to\psi_2$ to find
\bea
T_2 =\sum_{\psi_1\in S_{\vec p}\times S_{\vec m}}\sum_{\psi_2\in S_{\vec p'}\times S_{\vec m'}}
\langle\vec{p}',\vec{m}'|\psi_2^{-1}E^{\psi_1^{-1}(1)}_{ci}E^{\psi_1^{-1}(p+1)}_{kc}|\vec p,\vec m\rangle
\langle \vec p,\vec m|\sigma_1^{-1}E^{\psi_1^{-1}(1)}_{ak}E^{\psi_1^{-1}(p+1)}_{ia}
\psi_2\sigma_2 |\vec p',\vec m'\rangle
\nonumber
\eea
We will use $\hat b$ to denote the $q$ dimensional vector that has all entries zero except the $b$th entry which is 1. 
For a non-zero contribution the first factor requires
\bea
\vec p-\hat i+\hat c &=&\vec p'\cr
\vec m-\vec c+\vec k &=& \vec m'
\eea
and the second factor requires
\bea
\vec m-\hat i+\hat a &=&\vec m'\cr
\vec p-\vec a+\vec k &=& \vec p'
\eea
There are two solutions:

{\vskip 0.2cm}

\underline{Case 1:} $\hat c =\hat i$ and $\hat a = \hat k$. In this case $\vec p = \vec p'$  and $\vec m -\hat i + \hat k 
= \vec m'$.

\underline{Case 2:} $\hat c =\hat k$ and $\hat a = \hat i$. In this case $\vec m = \vec m'$  and $\vec p -\hat i + \hat k 
= \vec p'$.

{\vskip 0.2cm}

\noindent
For case 1
\bea
T_2 =\sum_{\psi_1\in S_{\vec p}\times S_{\vec m}}\sum_{\psi_2\in S_{\vec p'}\times S_{\vec m'}}
\langle\vec{p}',\vec{m}'|\psi_2^{-1}E^{\psi_1^{-1}(1)}_{ii}E^{\psi_1^{-1}(p+1)}_{ki}|\vec p,\vec m\rangle
\langle \vec p,\vec m|\sigma_1^{-1}E^{\psi_1^{-1}(1)}_{kk}E^{\psi_1^{-1}(p+1)}_{ik}
\psi_2\sigma_2 |\vec p',\vec m'\rangle
\nonumber
\eea
Consider the sum over $\psi_1$. Due to the factor $E^{\psi_1^{-1}(p+1)}_{ki}$ we get a non-zero contribution from 
the slots $p + 1, p + 2,\cdots , p + m$ (a $Y$ string) if a string starts from node $k$ and ends at node $i$. 
Thus, the sum over $\psi_1$ gives
\bea
T_2 &=& (p - 1)!(m - 1)!p_{i\to k}m_{ii}\sum_{\psi_2\in S_{\vec p}\times S_{\vec m'}}
\langle\vec{p},\vec{m}'|\psi_2^{-1}|\vec p,\vec m'\rangle
\langle \vec p,\vec m'|\sigma_1^{-1}\psi_2\sigma_2 |\vec p,\vec m'\rangle\cr
&=& (p - 1)!(m - 1)!p_{i\to k}m_{ii}\sum_{\psi_2\in S_{\vec p}\times S_{\vec m'}}\sum_{h_1,h_2\in H_X\times H_Y}
\delta (\psi_2^{-1}h_1)\delta (\sigma_1^{-1}\psi_2\sigma_2h_2)\cr
&=& (p - 1)!(m - 1)!p_{i\to k}m_{ii}\sum_{h_1,h_2\in H_X\times H_Y}
\delta (\sigma_1^{-1}h_1\sigma_2h_2)
\eea
For case 2
\bea
T_2 =\sum_{\psi_1\in S_{\vec p}\times S_{\vec m}}\sum_{\psi_2\in S_{\vec p'}\times S_{\vec m}}
\langle\vec{p}',\vec{m}|\psi_2^{-1}E^{\psi_1^{-1}(1)}_{ki}E^{\psi_1^{-1}(p+1)}_{kk}|\vec p,\vec m\rangle
\langle \vec p,\vec m|\sigma_1^{-1}E^{\psi_1^{-1}(1)}_{ik}E^{\psi_1^{-1}(p+1)}_{ii}
\psi_2\sigma_2 |\vec p',\vec m\rangle
\nonumber
\eea
Consider the sum over $\psi_1$. 
We get a non-zero contribution for each $Y$ string starting from node $k$ which ends at node $i$.
After summing over $\psi_1$ we have
\bea
T_2 &=& (p - 1)!(m - 1)!p_{ii}m_{k\to i}\sum_{\psi_2\in S_{\vec p'}\times S_{\vec m}}
\langle\vec{p}',\vec{m}|\psi_2^{-1}|\vec p,\vec m'\rangle
\langle \vec p',\vec m|\sigma_1^{-1}\psi_2\sigma_2 |\vec p',\vec m\rangle\cr
&=& (p - 1)!(m - 1)!p_{ii}m_{k\to i}\sum_{\psi_2\in S_{\vec p'}\times S_{\vec m}}\sum_{h_1,h_2\in H_X\times H_Y}
\delta (\psi_2^{-1}h_1)\delta (\sigma_1^{-1}\psi_2\sigma_2h_2)\cr
&=& (p - 1)!(m - 1)!p_{ii}m_{k\to i}\sum_{h_1,h_2\in H_X\times H_Y}
\delta (\sigma_1^{-1}h_1\sigma_2h_2)
\eea

Armed with these sums, we now obtain a rather explicit expression for the matrix elements of $D^{XY}_2$ in the Gauss 
graph basis
\bea
M^{\vec m,\vec p}_{R,r,\sigma_1\,\,T,t,\sigma_2}&=&-g_{YM}^2
{\delta_{ru}\over\sqrt{|O_{R,r}^{\vec m,\vec p}(\sigma_1)|^2 |O_{T,t}^{\vec m,\vec p}(\sigma_2)|^2}}
\sum_{R'}\delta_{R'_iT'_k}\sqrt{c_{RR'}c_{TT'}\over l_{R_i}l_{T_k}}\cr
&\times& [2\delta_{ik}p_i m_i - p_{ki}m_{ii}-  p_{ii}m_{ik}]
\sum_{h_1,h_2\in H_X\times H_Y}
\delta (\sigma_1^{-1}h_1\sigma_2h_2)\cr
&&\label{finalD}
\eea
This is the key result of this section and one of the key results of this paper.
We will now describe how the above matrix can be diagonalized.

\section{Boson Lattice}\label{blattice}

Our goal in this section is to diagonalize (\ref{finalD}). 
This is achieved by interpreting (\ref{finalD}) as the matrix elements of a Hamiltonian for bosons on a lattice.
Towards this end, first note that the matrix elements $M^{\vec m,\vec p}_{R,r,\sigma_1\,\,T,t,\sigma_2}$ are only
non-zero if we can choose coset representatives such that $\sigma_1$ and $\sigma_2$ describe the same element of 
$S_m\times S_p$. 
This implies that the brane-string systems described by $\sigma_1$ and $\sigma_2$ differ only in the number of strings 
with both ends attached to the same brane, but not in the number of string stretching between distinct branes. 
This already implies that the contribution $D^{XY}_2$ only mixes eigenstates of $D^{XZ}_2$ and $D^{YZ}_2$ that 
are degenerate and hence that all three are simultaneously diagonalizable. 
In this case the matrix element in (\ref{finalD}) simplifies to
\bea
M^{\vec m,\vec p}_{R,r,\sigma_1\,\,T,t,\sigma_2}&=&-g_{YM}^2
{\sqrt{|O_{R,r}^{\vec m,\vec p}(\sigma_1)|^2\over |O_{T,t}^{\vec m,\vec p}(\sigma_2)|^2}}
\delta_{ru}\delta_{R'_iT'_k}\sqrt{(N+l_{R_i})(N+l_{T_k})\over l_{R_i}l_{T_k}}\cr\cr
&\times& \Big[2\delta_{ik}p_i (\sigma_2)m_i(\sigma_2) - p_{ki}m_{ii}(\sigma_2)-  p_{ii}(\sigma_2)m_{ik}\Big]
\eea

The number of strings stretching between the branes $m_{ik}$ (for $Y$ strings) and $p_{ki}$ (for $X$ strings) are the 
same for both systems so that
\bea
m_{ik}(\sigma_1) = m_{ik}(\sigma_2) \equiv m_{ik}\qquad p_{ik}(\sigma_1) = p_{ik}(\sigma_2) \equiv p_{ik}
\eea
It is the number of closed loops ($m_{ii}$ for $Y$ loops and $p_{ii}$ for $X$ loops) that can differ between the operators 
that mix. 
Finally, we have introduced the notation
\bea
p_i(\sigma ) =\sum_{k\ne i} p_{ik} + p_{ii}(\sigma)\qquad
m_i(\sigma ) =\sum_{k\ne i} m_{ik} + m_{ii}(\sigma)
\eea
From the structure of the operator mixing problem, we would expect that
$M^{\vec m,\vec p}_{R,r,\sigma_1\,\,T,t,\sigma_2}=M^{\vec m,\vec p}_{T,t,\sigma_2\,\, R,r,\sigma_1}$. 
This is indeed the case, as a consequence of the easily checked identity
\bea
{\sqrt{|O_{R,r}^{\vec m,\vec p}(\sigma_1)|^2\over |O_{T,t}^{\vec m,\vec p}(\sigma_2)|^2}}
\Big[2\delta_{ik}p_i (\sigma_2)m_i(\sigma_2) - p_{ki}m_{ii}(\sigma_2)-  p_{ii}(\sigma_2)m_{ik}\Big]\cr
={\sqrt{|O_{T,t}^{\vec m,\vec p}(\sigma_2)|^2\over |O_{R,r}^{\vec m,\vec p}(\sigma_1)|^2}}
\Big[2\delta_{ik}p_i (\sigma_1)m_i(\sigma_1) - p_{ki}m_{ii}(\sigma_1)-  p_{ii}(\sigma_1)m_{ik}\Big]\cr
\eea
which holds for any $i, k$.

The lattice model consists of two distinct species of bosons, one for $X$ and one for $Y$, hopping on a lattice, with a site 
for every brane, or equivalently, a site for every row in the Young diagram $R$ labeling the Gauss graph operator 
$\hat O^{\vec m,\vec p}_{R,r}(\sigma)$.
The bosons are described by the following commuting sets of operators
\bea
\big[ a_i,a_j^\dagger\big]=\delta_{ij}\qquad
\big[ a_i^\dagger ,a_j^\dagger\big]=0=\big[ a_i,a_j\big]\cr\cr
\big[ b_i,b_j^\dagger\big]=\delta_{ij}\qquad
\big[ b_i^\dagger ,b_j^\dagger\big]=0=\big[ b_i,b_j\big]
\eea
Using these boson oscillators, we have
\bea
m_{ii} = a_i^\dagger a_i\qquad p_{ii} = b_i^\dagger b_i
\eea
\bea
m_i = \sum_k m_{ik}+a_i^\dagger a_i\qquad
p_i = \sum_k p_{ik}+b_i^\dagger b_i
\eea

The vacuum of the Fock space $|0\rangle$ obeys
\bea a_i|0\rangle =0=b_i|0\rangle\qquad i=1,2,\cdots ,q.
\eea
The Hamiltonian of the lattice model is given by
\bea
H &=&\sum_{i,j=1}^q\sqrt{(N+l_{R_i})(N+l_{R_j})\over l_{R_i}l_{R_j}}\Bigg(
2\delta_{ij}\Big(\sum_{l\ne i} p_{il} + b_i^\dagger b_i\Big)\Big(\sum_{l\ne i} m_{il} + a_i^\dagger a_i\Big)\cr
&&\qquad -p_{ji}a_j^\dagger a_i -m_{ji}b_j^\dagger b_i\Bigg)
\label{FDform}
\eea
Notice that this Hamiltonian is quadratic in each type of oscillator. It has a nontrivial repulsive interaction given by the
$\sum_i a_i^\dagger a_i b_i^\dagger b_i$  term, which makes it energetically unfavorable for $a$ and $b$ type particles 
to sit on the same site. Also, the full Fock space is a tensor product between the Fock space for the $a$ oscillator and the 
Fock space for the $b$ oscillator. We will use the occupation number representation to describe the boson states. 
To complete the mapping to the lattice model, we need to explain the correspondence between Gauss graph operators 
and states of the boson lattice. 
This map is given by reading the boson occupation numbers for each site from the number of closed strings with both 
ends attached to the node corresponding to that site. 
In the next subsection we consider an example which nicely illustrates this map.

Finally, lets make an important observation regarding (\ref{FDform}). 
Although the eigenvalues of this Hamiltonian are subleading contributions to the anomalous dimension, there is an 
important situation in which this correction is highly significant: for BPS states the leading contribution to the anomalous
dimension vanishes and this subleading correction is important. 
The BPS operators are labeled by Gauss graphs that have $p_{ik} = m_{ik} = 0$ whenever $i\ne k$, i.e. there are no 
strings stretching between branes. 
In this case, it is clear that (\ref{FDform}) vanishes so that the BPS operators remain BPS.

\subsection{Example}

\begin{figure}[h]
\begin{center}
\resizebox{!}{8cm}{\includegraphics{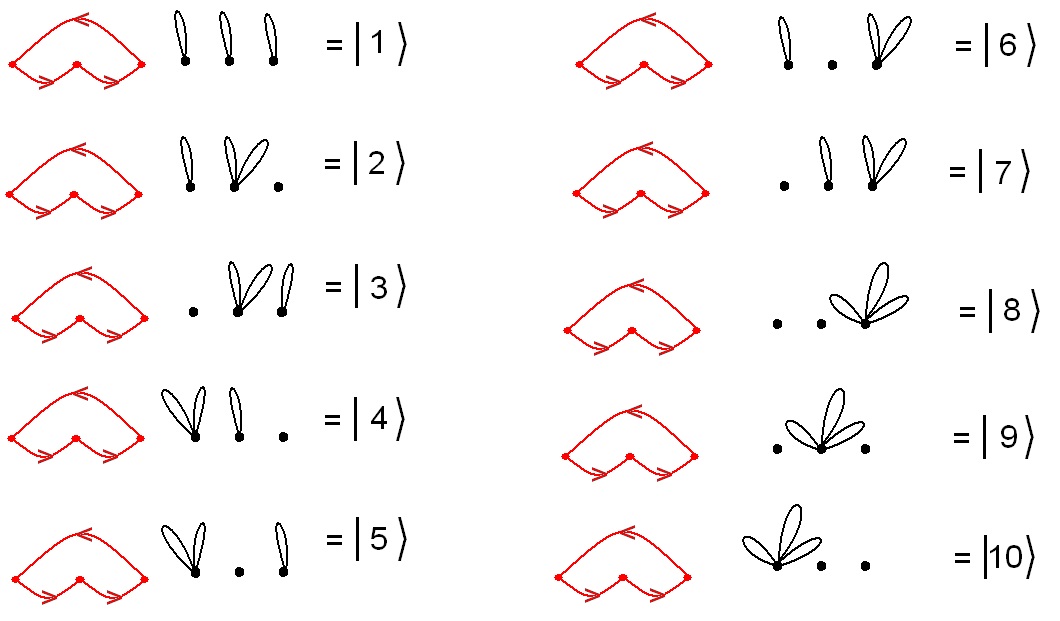}}
\caption{Each Gauss graph label is composed of two graphs, the first for the $X$ strings and the second for the 
$Y$ strings. Each graph has 3 nodes (because $q$ = 3). There are no $b$ type particles because there are no 
closed $X$ strings. There are 3 $a$ type particles because there are three closed $Y$ strings. All operators share 
the same $r$ label.}
 \label{NiceGG}
\end{center}
\end{figure}

In this section we will consider an example for which $R$ has $q = 3$ rows and $p = m = 3$. 
In this problem, 10 operators mix. 
The Gauss graph labels for the operators that mix are displayed in Figure \ref{NiceGG}.

For the Gauss graph operators shown, we have the following correspondence with boson lattice states
\bea
|1\rangle &=& a_1^\dagger a_2^\dagger a_3^\dagger |0\rangle \qquad 
|2\rangle = a_1^\dagger {(a_2^\dagger)^2\over\sqrt{2!}}|0\rangle\cr
|3\rangle &=& a_3^\dagger {(a_2^\dagger)^2\over\sqrt{2!}}|0\rangle\qquad
|4\rangle = a_2^\dagger {(a_1^\dagger)^2\over\sqrt{2!}}|0\rangle\cr
|5\rangle &=& a_3^\dagger {(a_1^\dagger)^2\over\sqrt{2!}}|0\rangle\qquad
|6\rangle = a_1^\dagger {(a_3^\dagger)^2\over\sqrt{2!}}|0\rangle\cr
|7\rangle &=& a_2^\dagger {(a_3^\dagger)^2\over\sqrt{2!}}|0\rangle\qquad
|8\rangle =  {(a_3^\dagger)^3\over\sqrt{3!}}|0\rangle\cr
|9\rangle &=&  {(a_2^\dagger)^3\over\sqrt{3!}}|0\rangle\qquad
|10\rangle =  {(a_1^\dagger)^3\over\sqrt{3!}}|0\rangle
\eea

It is now rather straight forwards to compute matrix elements of the lattice Hamiltonian. 
For example
\bea
\langle 1|H|2\rangle = -\sqrt{(N+l_{R_3})(N+l_{R_2})\over l_{R_2}l_{R_3}}\sqrt{2}
\eea
It is instructive to compare this to the answer coming from (\ref{finalD}). 
To move from state 2 to state 1, a string must detach from node 2 and reattach to node 3. 
Thus, we should plug $i = 2$ and $k = 1$ into (\ref{finalD}). 
The Gauss graph $\sigma_1$ corresponds to $|1\rangle$ while $\sigma_2$ corresponds to $|2\rangle$. 
In addition $R'_2=T'_1$ and from the Gauss graphs we read off $p_{32} = 1$ and $m_{22}(\sigma_1) = 1$. 
It is now simple to see that (\ref{finalD}) is in complete agreement with the above matrix element.

\begin{figure}[h]
\begin{center}
\resizebox{!}{6cm}{\includegraphics{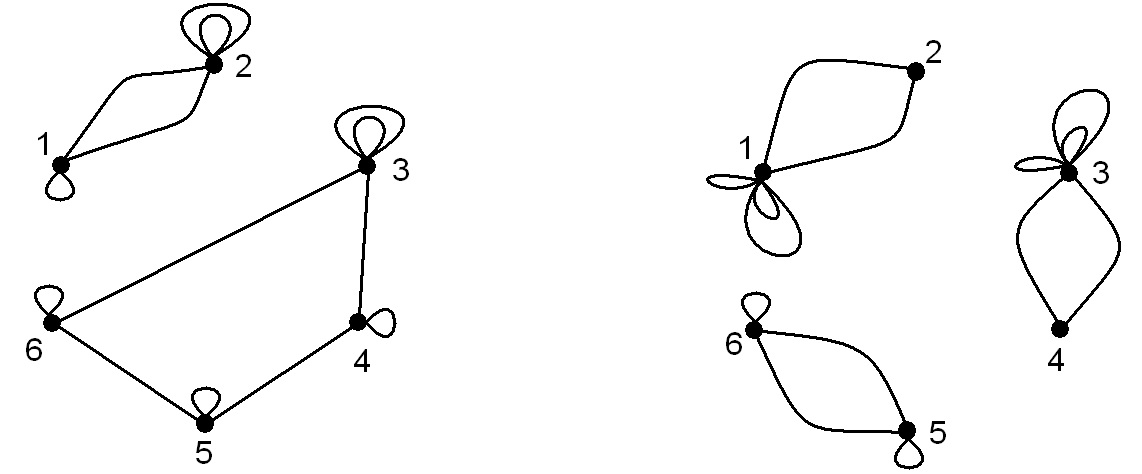}}
\caption{An example of a Gauss graph with non-zero a and b occupation numbers.}
 \label{biggauss}
\end{center}
\end{figure}

Finally, the state corresponding to the Gauss graph in Figure \ref{biggauss} is
\bea
{(a_1^\dagger)^3\over \sqrt{3!}}
{(a_3^\dagger)^3\over \sqrt{3!}}
a_5^\dagger a_6^\dagger b_1^\dagger
{(b_2^\dagger)^2\over \sqrt{2!}}
{(b_3^\dagger)^2\over \sqrt{2!}}
b_4^\dagger
b_5^\dagger
b_6^\dagger |0\rangle
\eea

\section{Diagonalization}\label{diag}

In this section we will consider a class of examples that can be diagonalized explicitly. 
Our main motivation is to show that working with the lattice is simple, so the mapping we have found is useful.

\subsection{Exact Eigenstates}

\begin{figure}[h]
\begin{center}
\resizebox{!}{4cm}{\includegraphics{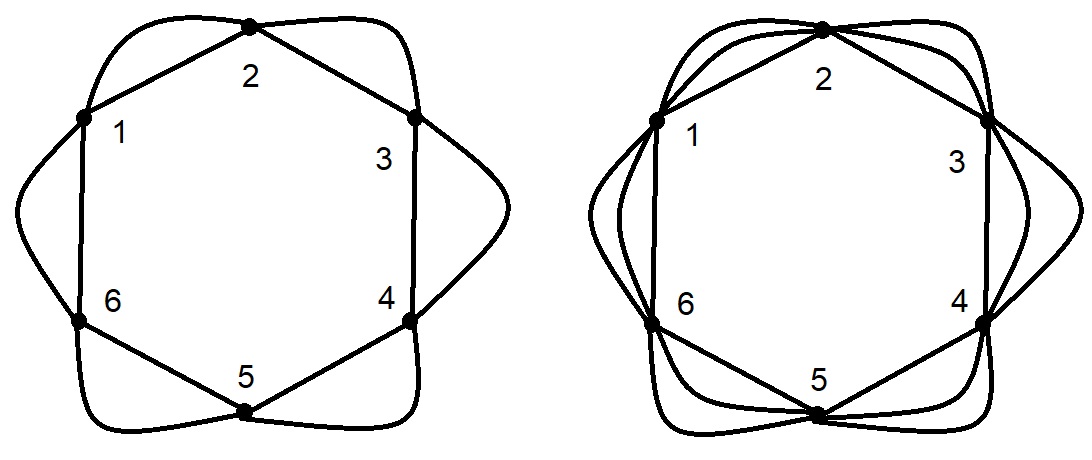}}
\caption{An example of a Gauss graph that is easily solvable. The example shown has $A = 2$ and $B = 3$.}
 \label{typGG}
\end{center}
\end{figure}

For these examples take
\bea
p_{ki} = p_{ik} = \delta_{k,i+1}B\qquad  m_{ki} = m_{ik} = \delta_{k,i+1}A 
\eea
with $A$ and $B$ two positive integers. 
For examples of Gauss graphs that obey this condition, see Figure \ref{typGG}. 
There are two cases we will consider: we will fix the number of $a$ particles to zero and leave the number of $b$ particles
arbitrary, or, fix the number of $b$ particles to zero and leave the number of $a$ particles arbitrary. 
We will also specialize to labels $R$ that have the difference between any two row lengths $l_{R_i} -l_{R_j}\sim N$, but
${l_{R_i}-l_{R_j}\over l_{R_i}}\approx 0$. 
In this case our lattice Hamiltonian simplifies to
\bea
H ={(N + l_{R_1})\over l_{R_1}}\sum_{i=1}^q\left( 2\left(B+b_i^\dagger b_i\right)\left(A+a_i^\dagger a_i\right)\right.\cr
-B(a_i^\dagger a_{i+1}+a_{i+1}^\dagger a_i)-A(b_i^\dagger b_{i+1}+b_{i+1}^\dagger b_i)\Big)
\eea
This Hamiltonian is easily diagonalized by going to Fourier space. 
Indeed, in terms of the new oscillators
\bea
\tilde a_n ={1\over\sqrt{q}}\sum_{k=1}^q e^{i{2\pi kn\over q}}a_k\qquad
\tilde b_n ={1\over\sqrt{q}}\sum_{k=1}^q e^{i{2\pi kn\over q}}b_k\qquad
n = 0, 1, \cdots  , q - 1 
\eea
the Hamiltonian becomes (we have set the number of $a$ particles to zero)
\bea
H = A{(N + l_{R_1})\over l_{R_1}}\sum_{n=0}^{q-1}\left(2-2\cos\left({2\pi n\over q}\right)\right)
\tilde b_n^\dagger \tilde b_n+2AB q {(N + l_{R_1})\over l_{R_1}}
\eea
Eigenstates of the lattice Hamiltonian are given by arbitrary momentum space excitations
\bea
\prod_{n=0}^{q-1} {(\tilde a_n^\dagger)^{\alpha_n}\over\sqrt{\alpha_n !}}|0\rangle\qquad 
{\rm or}\qquad \prod_{n=0}^{q-1} {(\tilde b_n^\dagger)^{\beta_n}\over\sqrt{\beta_n !}}|0\rangle
\eea
where the occupation numbers $\alpha_n,\beta_n$ are arbitrary. 
This state can be translated back into the Gauss graph language to give operators of a definite scaling dimension.

\subsection{General Properties of Low Energy Eigenstates}

In this section we will sketch the features of generic low energy states of the lattice Hamiltonian. 
We begin by relaxing the constraint that only one species is hopping. 
In the end we will also make comments valid for the general Gauss graph configuration. 
The Hamiltonian becomes
\bea
H = H_a + H_b + H_{ab} + E_0
\eea
\bea
H_a ={(N + l_{R_1})\over l_{R_1}} B\sum_{i=1}^q 
\left( 2a^\dagger_i a_i-a_i^\dagger a_{i+1}-a_{i+1}^\dagger a_i\right)
\eea

\bea
H_b ={(N + l_{R_1})\over l_{R_1}} A\sum_{i=1}^q 
\left( 2b^\dagger_i b_i-b_i^\dagger b_{i+1}- b_{i+1}^\dagger b_i\right)
\eea

\bea
H_{ab} ={(N + l_{R_1})\over l_{R_1}}\sum_{i=1}^q 2 b^\dagger_i b_i a^\dagger_i a_i
\eea
The constant $E_0 = 2ABq {(N+l_{R_1})\over l_{R_1}}$ is not important for the dynamics but must be included to obtain 
the correct anomalous dimensions. 
To start, consider $H_a$ which is a kinetic term for the $a$ particles. 
The first term in the Hamiltonian implies that it costs energy to have an $a$ particle occupying a site, while
the second and third terms tell us this energy can be lowered by hopping between sites $i$ and $i + 1$. 
Consequently, to minimize $H_a$, the $a$ particles will spread out as much as is possible. 
This is in perfect accord with the results of the last section. 
The lowest energy single particle state is the zero momentum state, which occupies each site with the same probability: the
particle spreads out as much as is possible. 
Very similar reasoning for $H_b$ implies that the $b$ particles will also spread out as much as is possible. 
Finally, the term $H_{ab}$ is a repulsive interaction, telling us that it costs energy to have $a$s and $b$s
occupying the same site. 
So there is a competition going on: The terms $H_a$ and $H_b$ want to spread the $a$s and $b$s uniformly on the 
lattice which would certainly distribute $a$s and $b$s to the same site. 
The term $H_{ab}$ wants to ensure that any particular site will have only $a$s or $b$s but not both. Who wins?

Consider a thermodynamic like limit where we consider a very large number of both species of particles, $n_a$ and $n_b$. 
In the end, the low energy state will be a ``demixed" state with no sites holding both $a$s and $b$s. 
To see this, note that $H_a$ grows like $n_a$ and $H_b$ like $n_b$. 
This is much smaller than the growth of the term $H_{ab}$ which grows like $n_a n_b$, so the repulsive interaction wins.
This conclusion is nicely borne out by numerical results for the two component Bose-Hubbard model\cite{bhm1,bhm2}. 
The ground state phase diagram of the Hamiltonian of \cite{bhm1}, shows four distinct phases: double super
fluid phase, supercounterflow phase, demixed Mott insulator phase and a demixed superfluid phase. 
Comparing our Hamiltonian to that of \cite{bhm1}, we are always in the demixed superfluid phase: the $a$ and $b$ 
particles do not mix, but are free to move in their respective domains.

For the generic Gauss graph, with any choices for the values of $m_{ik}$ and $p_{ik}$, it is clear that $H_a$ and $H_b$ 
will still cause the $a$ and $b$ particles to spread out as much as possible. 
The term $H_{ab}$ will again dominate when we have large numbers of $a$s and $b$s so we again expect a demixed gas. 
We can translate this structure of the generic state back into the language of the giant graviton description. 
Up to now we have considered dual giant gravitons which correspond to operators labeled by Young diagrams with long rows.
Recall that dual giant gravitons wrap an S$^3\subset$AdS$_5$. 
In this context, $l_{R_1}$ is the momentum of each giant and $N +l_{R_1}$ is the radius on the LLM plane at which
the giant orbits. The Hamiltonian for giant gravitons, which wrap S$^3\subset$S$^5$ is given by
\bea
H ={(N - l_{R_1})\over l_{R_1}}\sum_{i=1}^q\left( 2\left(B+b_i^\dagger b_i\right)\left(A+a_i^\dagger a_i\right)\right.\cr
-B(a_i^\dagger a_{i+1}+a_{i+1}^\dagger a_i)-A(b_i^\dagger b_{i+1}+b_{i+1}^\dagger b_i)\Big)
\eea

These operators are labeled by Young diagrams with long columns. 
The giants orbit on the LLM plane with a radius of $N- l_{R_1}$. 
The $X$ and $Y$ fields are each charged under different $U(1)$s of the ${\cal R}$-symmetry group. 
The ${\cal R}$-symmetry of the CFT translates into angular momentum of the dual string theory, so that attaching the 
particles to a given giant corresponds to giving the giant angular momentum. 
The lowest energy giant graviton states are obtained by distributing the momenta carried by the $X$ and $Y$ fields 
evenly between the giants with the condition that any particular giant carries only $X$ or $Y$ momenta, but not both. 
These conclusions hold for the generic state where there are enough $p_{ik}$ and $m_{ik}$ non-zero, allowing the $X$s and
$Y$s to hop between any two giants, possibly by a complicated path. 
Thus in the end we see that the mapping to the boson lattice model has allowed a rather
detailed understanding of the operator mixing problem.

\section{Conclusions}\label{conc}

In this article we have studied the action of the one loop dilatation operator $D_2$ on Gauss graph operators 
$O^{\vec m,\vec p}_{R,r} (\sigma)$ which belong to the $SU(3)$ sector.
The term we have studied, $D^{XY}_2$ , is diagonal in the $r$ label, mixing operators labeled by distinct graphs. 
It makes a subleading contribution as compared to $D^{XZ}_2$ and $D^{YZ}_2$ when $n\gg m + p$. 
The two leading terms mix operators labeled by distinct $r$s. 
Diagonalizing the action of $D^{XZ}_2$ and $D^{YZ}_2$ on $r$ leads to a collection of decoupled harmonic 
oscillators, which we refer to as the $Z$ oscillators, since the $r$ label is associated with $Z$. 
The spectrum of the $Z$ oscillators gives the leading contribution to the anomalous dimensions. 
The new contribution that we have studied in this paper can also be mapped to a collection of oscillators, describing a 
lattice boson model. 
This is done by introducing two sets of oscillators, the $X$ and $Y$ oscillators associated to the $X$ and $Y$ fields.
Diagonalizing the $X$ and $Y$ oscillators breaks degeneracies among different copies of $Z$ oscillators and leads to a 
constant addition to their ground state energy. 
This is then a constant shift of the anomalous dimension. 
Although this shift is subleading (it is of order ${m\over n}$), it could potentially show that certain states are not in fact BPS. This was investigated in detail and it turns out that states that are BPS (their leading order anomalous dimension vanishes) 
at leading order, remain BPS when the subleading correction is computed (it too vanishes).

The mapping that we have found to a lattice boson model has achieved an enormous simplification of the operator mixing problem and we have managed to understand it in some detail. 
Indeed, using the lattice boson model, we have argued that the lowest energy giant graviton states are obtained by
distributing the momenta carried by the $X$ and $Y$ fields evenly between the giants with the condition that any 
particular giant carries only $X$ or $Y$ momenta, but not both.
Since states with two charges are typically $1/4$-BPS while states with 3 charges are typically $1/8$-BPS, it maybe that
the solution is locally trying to maximize susy.  
It would be interesting to arrive at the same picture, employing the dual string theory description.

Perhaps the most interesting consequence of our results is that they suggest ways in which one can go beyond the 1/2 
BPS sector. 
Indeed, all three types of fields considered have been mapped to oscillators, so perhaps there is a more general description 
of this sector that treats all three types of oscillators on the same footing. 
This would relax the constraint $n\gg p + m$ which allows for operators that are far from the 1/2 BPS limit. 
Deriving this picture is a fascinating open problem, since it will require that we go beyond the displaced corners 
approximation, or alternatively, that we generalize it.

As a final comment, recall that Mikhailov \cite{Mikhailov:2000ya} has constructed an infinite family of $1/8$ BPS giant 
graviton branes in AdS$_5\times$S$^5$. 
Quantizing the space of Mikhailov's solutions leads to $N$ non-interacting bosons in a harmonic
oscillator\cite{Beasley:2002xv,Biswas:2006tj,Mandal:2006tk}. 
It is tempting to speculate that it is precisely these oscillators that we are uncovering in our study; for evidence in 
harmony with this suggestion see \cite{deMelloKoch:2017ytj}. It would be interesting to make this speculation precise.

\bigskip 

\begin{center} 
{ \bf Acknowledgements}
\end{center} 

We would like to thank Vishnu Jejjala, Sanjaye Ramgoolam and Izak Snyman for helpful
discussions. This work is supported by the South African Research Chairs
Initiative of the Department of Science and Technology and National Research
Foundation as well as funds received from the National Institute for
Theoretical Physics (NITheP).


\vskip3cm

\begin{thebibliography}{} 

\bibitem{Maldacena:1997re} 
  J.~M.~Maldacena,
  ``The Large N limit of superconformal field theories and supergravity,''
  Int.\ J.\ Theor.\ Phys.\  {\bf 38}, 1113 (1999)
  [Adv.\ Theor.\ Math.\ Phys.\  {\bf 2}, 231 (1998)]
  [hep-th/9711200].

\bibitem{Gubser:1998bc} 
  S.~S.~Gubser, I.~R.~Klebanov and A.~M.~Polyakov,
  ``Gauge theory correlators from noncritical string theory,''
  Phys.\ Lett.\ B {\bf 428}, 105 (1998)
  [hep-th/9802109].

\bibitem{Witten:1998qj} 
  E.~Witten,
  ``Anti-de Sitter space and holography,''
  Adv.\ Theor.\ Math.\ Phys.\  {\bf 2}, 253 (1998)
  [hep-th/9802150].

\bibitem{Gromov:2013pga} 
  N.~Gromov, V.~Kazakov, S.~Leurent and D.~Volin,
  ``Quantum Spectral Curve for Planar $\mathcal{N} = 4$ Super-Yang-Mills Theory,''
  Phys.\ Rev.\ Lett.\  {\bf 112}, no. 1, 011602 (2014)
  [arXiv:1305.1939 [hep-th]].

\bibitem{Minahan:2002ve} 
  J.~A.~Minahan and K.~Zarembo,
  ``The Bethe ansatz for N=4 superYang-Mills,''
  JHEP {\bf 0303}, 013 (2003)
  [hep-th/0212208].

\bibitem{Beisert:2010jr} 
  N.~Beisert {\it et al.},
  ``Review of AdS/CFT Integrability: An Overview,''
  Lett.\ Math.\ Phys.\  {\bf 99}, 3 (2012)
  [arXiv:1012.3982 [hep-th]].

\bibitem{Beisert:2005bm} 
  N.~Beisert, V.~A.~Kazakov, K.~Sakai and K.~Zarembo,
  ``The Algebraic curve of classical superstrings on AdS(5) x S**5,''
  Commun.\ Math.\ Phys.\  {\bf 263}, 659 (2006)
  [hep-th/0502226].

\bibitem{Lin:2004nb} 
  H.~Lin, O.~Lunin and J.~M.~Maldacena,
  ``Bubbling AdS space and 1/2 BPS geometries,''
  JHEP {\bf 0410}, 025 (2004)
  [hep-th/0409174].

\bibitem{McGreevy:2000cw} 
  J.~McGreevy, L.~Susskind and N.~Toumbas,
  ``Invasion of the giant gravitons from Anti-de Sitter space,''
  JHEP {\bf 0006}, 008 (2000)
  [hep-th/0003075].

\bibitem{Grisaru:2000zn} 
  M.~T.~Grisaru, R.~C.~Myers and O.~Tafjord,
  ``SUSY and goliath,''
  JHEP {\bf 0008}, 040 (2000)
  [hep-th/0008015].

\bibitem{Hashimoto:2000zp} 
  A.~Hashimoto, S.~Hirano and N.~Itzhaki,
  ``Large branes in AdS and their field theory dual,''
  JHEP {\bf 0008}, 051 (2000)
  [hep-th/0008016].

\bibitem{Balasubramanian:2001nh} 
  V.~Balasubramanian, M.~Berkooz, A.~Naqvi and M.~J.~Strassler,
  ``Giant gravitons in conformal field theory,''
  JHEP {\bf 0204}, 034 (2002)
  [hep-th/0107119].

\bibitem{Berenstein:2003ah} 
  D.~Berenstein,
  ``Shape and holography: Studies of dual operators to giant gravitons,''
  Nucl.\ Phys.\ B {\bf 675}, 179 (2003)
  [hep-th/0306090].

\bibitem{Corley:2001zk} 
  S.~Corley, A.~Jevicki and S.~Ramgoolam,
  ``Exact correlators of giant gravitons from dual N=4 SYM theory,''
  Adv.\ Theor.\ Math.\ Phys.\  {\bf 5}, 809 (2002)
  [hep-th/0111222].

\bibitem{deMelloKoch:2007rqf} 
  R.~de Mello Koch, J.~Smolic and M.~Smolic,
  ``Giant Gravitons - with Strings Attached (I),''
  JHEP {\bf 0706}, 074 (2007)
  [hep-th/0701066].

\bibitem{Kimura:2007wy} 
  Y.~Kimura and S.~Ramgoolam,
  ``Branes, anti-branes and brauer algebras in gauge-gravity duality,''
  JHEP {\bf 0711}, 078 (2007)
  [arXiv:0709.2158 [hep-th]].

\bibitem{Brown:2007xh} 
  T.~W.~Brown, P.~J.~Heslop and S.~Ramgoolam,
  ``Diagonal multi-matrix correlators and BPS operators in N=4 SYM,''
  JHEP {\bf 0802}, 030 (2008)
   [arXiv:0711.0176 [hep-th]].

\bibitem{Bhattacharyya:2008rb} 
  R.~Bhattacharyya, S.~Collins and R.~de Mello Koch,
  ``Exact Multi-Matrix Correlators,''
  JHEP {\bf 0803}, 044 (2008)
  [arXiv:0801.2061 [hep-th]].

\bibitem{Brown:2008ij} 
  T.~W.~Brown, P.~J.~Heslop and S.~Ramgoolam,
  ``Diagonal free field matrix correlators, global symmetries and giant gravitons,''
  JHEP {\bf 0904}, 089 (2009)
  [arXiv:0806.1911 [hep-th]].

\bibitem{Kimura:2008ac} 
  Y.~Kimura and S.~Ramgoolam,
  ``Enhanced symmetries of gauge theory and resolving the spectrum of local operators,''
  Phys.\ Rev.\ D {\bf 78}, 126003 (2008)
   [arXiv:0807.3696 [hep-th]].

\bibitem{Kimura:2012hp} 
  Y.~Kimura,
  ``Correlation functions and representation bases in free N=4 Super Yang-Mills,''
  Nucl.\ Phys.\ B {\bf 865}, 568 (2012)
  [arXiv:1206.4844 [hep-th]].

\bibitem{deMelloKoch:2007nbd} 
  R.~de Mello Koch, J.~Smolic and M.~Smolic,
  ``Giant Gravitons - with Strings Attached (II),''
  JHEP {\bf 0709}, 049 (2007)
  [hep-th/0701067].

\bibitem{Bekker:2007ea} 
  D.~Bekker, R.~de Mello Koch and M.~Stephanou,
  ``Giant Gravitons - with Strings Attached. III.,''
  JHEP {\bf 0802}, 029 (2008)
  [arXiv:0710.5372 [hep-th]].

\bibitem{Brown:2008rs} 
  T.~W.~Brown,
  ``Permutations and the Loop,''
  JHEP {\bf 0806}, 008 (2008)
  [arXiv:0801.2094 [hep-th]].

\bibitem{Koch:2010gp} 
  R.~de Mello Koch, G.~Mashile and N.~Park,
  ``Emergent Threebrane Lattices,''
  Phys.\ Rev.\ D {\bf 81}, 106009 (2010)
  [arXiv:1004.1108 [hep-th]].

\bibitem{DeComarmond:2010ie} 
  V.~De Comarmond, R.~de Mello Koch and K.~Jefferies,
  ``Surprisingly Simple Spectra,''
  JHEP {\bf 1102}, 006 (2011)
  [arXiv:1012.3884 [hep-th]].

\bibitem{Carlson:2011hy} 
  W.~Carlson, R.~de Mello Koch and H.~Lin,
  ``Nonplanar Integrability,''
  JHEP {\bf 1103}, 105 (2011)
  [arXiv:1101.5404 [hep-th]].

\bibitem{Koch:2011hb} 
  R.~de Mello Koch, M.~Dessein, D.~Giataganas and C.~Mathwin,
  ``Giant Graviton Oscillators,''
  JHEP {\bf 1110}, 009 (2011)
  [arXiv:1108.2761 [hep-th]].

\bibitem{deMelloKoch:2011ci} 
  R.~de Mello Koch, G.~Kemp and S.~Smith,
  ``From Large N Nonplanar Anomalous Dimensions to Open Spring Theory,''
  Phys.\ Lett.\ B {\bf 711}, 398 (2012)
  [arXiv:1111.1058 [hep-th]].

\bibitem{deMelloKoch:2012ck} 
  R.~de Mello Koch and S.~Ramgoolam,
  ``A double coset ansatz for integrability in AdS/CFT,''
  JHEP {\bf 1206}, 083 (2012)
  [arXiv:1204.2153 [hep-th]].

\bibitem{Koch:2015pga} 
  R.~de Mello Koch, N.~H.~Tahiridimbisoa and C.~Mathwin,
  ``Anomalous Dimensions of Heavy Operators from Magnon Energies,''
  JHEP {\bf 1603}, 156 (2016)
  [arXiv:1506.05224 [hep-th]].

\bibitem{Koch:2016jnm} 
  R.~de Mello Koch, C.~Mathwin and H.~J.~R.~van Zyl,
  ``LLM Magnons,''
  JHEP {\bf 1603}, 110 (2016)
  [arXiv:1601.06914 [hep-th]].

\bibitem{Beisert:2005tm} 
  N.~Beisert,
  ``The SU(2|2) dynamic S-matrix,''
  Adv.\ Theor.\ Math.\ Phys.\  {\bf 12}, 945 (2008)
  [hep-th/0511082].

\bibitem{Koch:2013yaa} 
  R.~de Mello Koch, S.~Graham and W.~Mabanga,
  ``Subleading corrections to the Double Coset Ansatz preserve integrability,''
  JHEP {\bf 1402}, 079 (2014)
  [arXiv:1312.6230 [hep-th]].

\bibitem{Balasubramanian:2004nb} 
  V.~Balasubramanian, D.~Berenstein, B.~Feng and M.~x.~Huang,
  ``D-branes in Yang-Mills theory and emergent gauge symmetry,''
  JHEP {\bf 0503}, 006 (2005)
  [hep-th/0411205].

\bibitem{bhm1}
F. Lingua, M. Guglielmino, V. Penna and B. Capogrosso Sansone,
``Demixing effects in mixtures of two bosonic species,'' Phys. Rev. {\bf A92}, 053610 (2015).

\bibitem{bhm2}
T. Mishra, R.V. Pai and B.P. Das, ``Phase separation in a two-species Bose mixture,'' Phys. Rev. {\bf A76}
013604 (2007).

\bibitem{Mikhailov:2000ya} 
  A.~Mikhailov,
  ``Giant gravitons from holomorphic surfaces,''
  JHEP {\bf 0011}, 027 (2000)
  [hep-th/0010206].

\bibitem{Beasley:2002xv} 
  C.~E.~Beasley,
  ``BPS branes from baryons,''
  JHEP {\bf 0211}, 015 (2002)
  [hep-th/0207125].

\bibitem{Biswas:2006tj} 
  I.~Biswas, D.~Gaiotto, S.~Lahiri and S.~Minwalla,
  ``Supersymmetric states of N=4 Yang-Mills from giant gravitons,''
  JHEP {\bf 0712}, 006 (2007)
  [hep-th/0606087].

\bibitem{Mandal:2006tk} 
  G.~Mandal and N.~V.~Suryanarayana,
  ``Counting 1/8-BPS dual-giants,''
  JHEP {\bf 0703}, 031 (2007)
  [hep-th/0606088].

\bibitem{deMelloKoch:2017ytj} 
  R.~de Mello Koch and L.~Nkumane,
  ``From Gauss Graphs to Giants,''
  arXiv:1710.09063 [hep-th].

\end{thebibliography}
\end{document}